\documentclass[10pt,letterpaper]{article}
\usepackage[top=0.85in,left=2.75in,footskip=0.75in]{geometry}

\usepackage{amsmath,amssymb}
\usepackage{changepage}
\usepackage{textcomp,marvosym}
\usepackage{cite}
\usepackage{nameref,hyperref}
\usepackage[right]{lineno}
\usepackage[nopatch=eqnum]{microtype}
\DisableLigatures[f]{encoding = *, family = * }
\usepackage[table]{xcolor}
\usepackage{array}

\usepackage{algorithm}
\usepackage{algpseudocode}
\usepackage{amsmath}

\DeclareMathOperator*{\argmin}{arg\,min}
\newcommand{\sampling}{\mathop{\sim}\limits}

\newcolumntype{+}{!{\vrule width 2pt}}
\newlength\savedwidth

\raggedright
\usepackage[aboveskip=1pt,labelfont=bf,labelsep=period,justification=raggedright,singlelinecheck=off]{caption}

\makeatletter
\renewcommand{\@biblabel}[1]{\quad#1.}
\makeatother

\usepackage{lastpage,fancyhdr,graphicx}
\usepackage{epstopdf}
\fancyheadoffset[L]{2.25in}
\fancyfootoffset[L]{2.25in}
\begin{document}
\vspace*{0.2in}

\begin{flushleft}
{\Large
\textbf\newline{Relaxation-assisted reverse annealing on nonnegative/binary matrix factorization
 } 
}
\newline
\\
Renichiro Haba\textsuperscript{1,2, *},
Masayuki Ohzeki\textsuperscript{1, 2, 3},
Kazuyuki Tanaka\textsuperscript{1},

\bigskip
\textbf{1} Graduate School of Information Sciences, Tohoku University, Sendai, Japan
\\
\textbf{2} Sigma-i Co., Ltd., Tokyo, Japan
\\
\textbf{3} Department of Physics, Institute of Science Tokyo, Tokyo, Japan
\\
\bigskip

%
%





* renichiro.haba.r6@dc.tohoku.ac.jp

\end{flushleft}
\section*{Abstract}
Quantum annealing has garnered significant attention as meta-heuristics inspired by quantum physics for combinatorial optimization problems. 
Among its many applications, nonnegative/binary matrix factorization stands out for its complexity and relevance in unsupervised machine learning.
The use of reverse annealing, a derivative procedure of quantum annealing to prioritize the search in a vicinity under a given initial state, helps improve its optimization performance in matrix factorization.
This study proposes an improved strategy that integrates reverse annealing with a linear programming relaxation technique.
Using relaxed solutions as the initial configuration for reverse annealing, we demonstrate improvements in optimization performance comparable to the exact optimization methods.
Our experiments on facial image datasets show that our method provides better convergence than known reverse annealing methods.
Furthermore, we investigate the effectiveness of relaxation-based initialization methods on randomized datasets, demonstrating a relationship between the relaxed solution and the optimal solution.
This research underscores the potential of combining reverse annealing and classical optimization strategies to enhance optimization performance.



\section*{Introduction}
In recent years, advancements in information and communication technologies have made it easier than ever to quantify and utilize vast amounts of data. 
This has led to a growing interest in addressing societal challenges through the effective use of such data. 
Optimization plays a crucial role in these efforts. 
It focuses on minimizing a given objective function within a set of constraints, often modeled mathematically. 

The inherent complexity of optimization problems has led to the development of new computing architectures that can handle large-scale combinatorial optimization problems.
Quantum annealing, a meta-heuristic inspired by quantum physics, has emerged as a promising approach to solving such problems \cite{kadowakiQuantumAnnealingTransverse1998a}.
Quantum annealing leverages quantum fluctuations to escape local minima and find the global minimum of a given objective function.
The method is specialized in solving quadratic unconstrained binary optimization (QUBO) problems, and many well-known problems can be translated into \cite{lucasIsingFormulationsMany2014b}.
In the ideal procedure, quantum annealing outputs the optimal solution by slowly decreasing the strength of the fluctuation. 
The quantum adiabatic theorem ensures that the ground state, which corresponds to the optimal solution, is obtained by evolving the system adiabatically \cite{suzukiResidualEnergiesSlow2005a, morita_mathematical_2008, ohzekiQuantumAnnealingIntroduction2011a}.
The hardware implementing quantum annealing developed by D-Wave Systems, Inc., has become commercially available, marking a significant milestone in its practical applications.
In contrast to theoretical aspects, the machines do not perform quantum annealing ideally, and their optimization performance is quite limited at the present stage.
However, the rapidness of sampling can be effective for attaining relatively good solutions as a heuristic solver.
To inspect the industry applicability of D-Wave's quantum annealer, a wide variety of practical use cases have been explored in finance\cite{rosenbergSolvingOptimalTrading2016a, orusForecastingFinancialCrashes2019,venturelliReverseQuantumAnnealing2019}, traffic\cite{neukartTrafficFlowOptimization2017b, hussainOptimalControlTraffic2020a, shikanaiTrafficSignalOptimization2023,habaRoutingSchedulingOptimization2024}, logistics\cite{feldHybridSolutionMethod2019a}, manufacturing\cite{ohzekiControlAutomatedGuided2019, habaTravelTimeOptimization2022a, venturelliQuantumAnnealingImplementation2016a}
, and marketing\cite{nishimuraItemListingOptimization2019}, as well as in decoding problems\cite{ideMaximumLikelihoodChannel2020, araiMeanFieldAnalysis2021b}.

In recent years, reverse annealing (RA) has been proposed as a derivative procedure of quantum annealing that prioritizes in local search\cite{chancellorModernizingQuantumAnnealing2017b, ohkuwaReverseAnnealingFully2018b, yamashiroDynamicsReverseAnnealing2019, d-wavesystemsReverseQuantumAnnealing}.
For a given certain classical state at the initial stage, RA searches in its vicinity and possibly find a non-trivial, lower-energy state nearby.
Unlike the standard quantum annealing, where the search begins with strong quantum fluctuations starting in a superposition of all possible weights, RA starts from a pre-determined classical state with zero transverse fields.
This paper refers to the standard quantum annealing as forward annealing (FA).
RA is integrated in D-Wave's quantum annealer, and it is effective in solving various optimization problems\cite{goldenReverseAnnealingNonnegative2021b, ikedaApplicationQuantumAnnealing2019, venturelliReverseQuantumAnnealing2019, habaTravelTimeOptimization2022a}.
The performance of RA is highly dependent on the property of the initial state, which can be obtained by any method, such as FA, classical algorithms, or random sampling.
In many cases, RA is placed as a post-processing step after FA by making up for the shortcomings in the result of FA.
However, recent studies have shown the impact of using RA and classical algorithms to leverage mutual forte.
For some problems, the effectiveness of RA utilizing initial states obtained by greedy algorithms has been reported\cite{venturelliReverseQuantumAnnealing2019,habaTravelTimeOptimization2022a}.

Among many applications of quantum annealing, machine learning is one of the promising fields where quantum annealing can be applied.
Nonnegative matrix factorization (NMF) has been widely used in unsupervised machine learning tasks, such as clustering, dimension reduction, and feature extraction\cite{leeLearningPartsObjects1999,leeAlgorithmsNonnegativeMatrix2001}.
NMF factorizes a given matrix into two nonnegative matrices, which can be interpreted as the basis and the coefficient matrices.
The coefficients represent the contribution of each basis to the original matrix.
Recent studies have shown that NMF can be constrained to binarized 
basis selections to enhance the contrast of feature selection\cite{omalleyNonnegativeBinaryMatrix2018b}.
This method is called nonnegative/binary matrix factorization (NBMF).
The binary constraint makes the problem more challenging as the search space becomes discrete.
As quantum annealing is specialized in solving binary optimization problems, NBMF is a suitable application for quantum annealing\cite{ottavianiLowRankNonNegative2018, asaokaNonnegativeBinaryMatrix2023}.
However, as the current D-Wave's quantum annealer is still developing and the performance is limited, the accuracy of the learning result is not always satisfactory, especially for large-scale problems.
For this reason, employing RA to enhance the performance of quantum annealing in NBMF has been proposed\cite{goldenReverseAnnealingNonnegative2021b}.
The preceding study has shown that RA can improve the performance of quantum annealing in NBMF over FA alone.
Generally, for NMF or NBMF, the alternating least squares (ALS) algorithm is used to iteratively reduce the errors between the original matrix and the product of the factorized matrices.
Similarly to the approach to use RA in the literature, the states in each iteration in the ALS algorithm were utilized as the initial configuration for RA.
This approach is effective in improving the performance of quantum annealing in NBMF.
However, the performance of RA still has room for improvement, and the error convergence of RA is inferior to the cases where the problems are solved optimally.
Specifically, the methods for initial states should be explored more to understand the performance limitations of RA.

This study proposes an improved strategy integrating RA with a linear programming relaxation technique.
Relaxation is a strategy for mapping an original hard problem to another one with a relaxed constraint that is easier to solve.
Its solution can be used to obtain information about the original problem.
For instance, in the context of NBMF, the binary constraints can be relaxed to linear constraints, which makes the problem capable of being solved by gradient-based optimization methods.
Specifically, we use a projected gradient descent method to solve the relaxed problem\cite{linProjectedGradientMethods2007}.
For applying RA, rounding the relaxed solution to binary is a natural choice for initial states, which gives us a good approximation of the original problem.
Although hybrid strategies of RA with initial states obtained by classical algorithms have been proposed in the literature, the use of relaxation has not been explored yet. In this paper, we will unveil the potential of relaxation-assisted RA.

The remainder of this paper is organized as follows.
In the next section, we briefly introduce NBMF and RA and describe our method of employing relaxation-assisted RA.
In the following section, we evaluate the proposed method's performance on facial image datasets traditionally used in NMF and NBMF.
We compare the results with the conventional methods, specifically, those obtained by FA and RA with initial states obtained by ALS for comparison in terms of convergence and accuracy.
Furthermore, we investigate the effectiveness of relaxation-based initialization methods on randomized datasets to understand the relationship between the relaxed solution and the optimal solution.
Finally, we discuss the results and the potential of relaxation-assisted RA in the context of quantum annealing for NBMF and other optimization problems.

\section*{Methods}
This section introduces NMF and NBMF and describes our method to integrate RA with linear programming relaxation.

\subsection*{Nonnegative Matrix Factorization}
Nonnegative matrix factorization (NMF) is a method to factorize a given matrix $V \in \mathbb{R}^{+m \times n}$ into two nonnegative matrices $W \in \mathbb{R}^{+m \times k}$ and $H \in \mathbb{R}^{+k \times n}$, where $k$ is the rank of the factorization and $\mathbb{R^{+}}$ denotes the set of nonnegative real numbers.
The matrix $W$ represents the basis matrix and contains the potential features of the original matrix, while the matrix $H$ represents the coefficient matrix and contains the contribution of each basis to the original matrix. 
In NMF, $W$ and $H$ are approximated by minimizing the Frobenius norm of the difference between the original matrix $V$ and the product of $W$ and $H$ as follows:
\begin{equation}
  \begin{split}
  \label{eq:nmf}
  \min_{W, H} & \ ||V - WH||_F^2 \\
  \text{subject to} \quad & W \in \mathbb{R}^{+m \times k} \\
  & H \in \mathbb{R}^{+k \times n},
  \end{split}
\end{equation}
where $||\cdot||_F$ denotes the Frobenius norm.
The problem is generally a non-convex and NP-hard problem \cite{vavasisComplexityNonnegativeMatrix2007}. 
This complexity means that finding an exact solution is computationally hard for large matrices, as the solution space is vast and contains multiple local minima. 
To tackle this challenge, NMF is typically solved using iterative algorithms that approximate a solution rather than solving it exactly. 
Among these methods, popular choices include alternating least squares (ALS) and multiplicative update rules \cite{vanbenthemFastAlgorithmSolution2004a, leeAlgorithmsNonnegativeMatrix2001}.

In the ALS algorithm, $W$ and $H$ are approximated by iteratively updating $W$ and $H$ to decrease the squared error as depicted in Algorithm \ref{alg:ALS_NMF}.
\begin{algorithm}[tbp]
  \caption{Alternating Least Squares for NMF}
  \label{alg:ALS_NMF}
  \begin{algorithmic}
      \renewcommand{\algorithmicrequire}{\textbf{Input:}}
      \renewcommand{\algorithmicensure}{\textbf{Output:}}
      \Require{A nonnegative matrix $V \in \mathbb{R}^{+m \times n}$, the rank $k$}
      \Ensure{Nonnegative and binary matrices $W \in \mathbb{R}^{+m \times k}$ and $H \in \mathbb{R}^{+k \times n}$}
      \State Initialize $W^1 \in \mathbb{R}^{+m \times k}$ and $H^1 \in \{0, 1\}^{k \times n}$.
      \While{not converged}
      \State  $W^{t+1}$ := $\argmin_{W \in \mathbb{R}^{+m \times k}} ||V - W H^t||_F^2$
      \State  $H^{t+1}$ := $\argmin_{H \in \mathbb{R}^{+k \times n}} ||V - W^{t+1} H||_F^2$
      \State $t := t + 1$
      \EndWhile
  \end{algorithmic}
\end{algorithm}
In ALS, it is important to solve each subproblem efficiently.
The method to solve the projected gradient descent (PGD) method is known to be effective for the nonnegative constraint\cite{linProjectedGradientMethods2007}.
In addition, PGD is commonly used for constrained optimization problems, where the solution is projected onto the feasible region in each iteration.
Specifically, the problem is generally expressed as:
\begin{equation}
  \begin{split}
  \label{eq:constrained_optimization}
  \min_{\mathbf{x} \in \mathbb{R}^n } & \ f(\mathbf{x}) \\
  \text{subject to} \quad & l_i \leq x_i \leq u_i, \quad i = 1, \ldots, n,
  \end{split}
\end{equation}
where $f$ is a continuous and differentiable function, and $l_i$ and $u_i$ are the lower and upper bounds of the variables, respectively.
In PGD, the states are updated by the following rule:
\begin{equation}
  \label{eq:pgd}
  \mathbf{x}' = P_{\mathbf{l}, \mathbf{u}} \left( \mathbf{x} - \alpha \nabla f(\mathbf{x}) \right),
\end{equation}
where $\alpha$ is the step size and  $P_{\mathbf{l}, \mathbf{u}}$ denotes the projection onto the feasible region defined as:
\begin{equation}
  \label{eq:projection}
  P_{\mathbf{l}, \mathbf{u}}(\mathbf{x}) = \max(\mathbf{l}, \min(\mathbf{u}, \mathbf{x})).
\end{equation}
The step size $\alpha$ is determined by the Almijo rule along the projection arc, and it realizes fast convergence \cite{bertsekasNonlinearProgramming1999}.
In this study, we do not describe the details of the Almijo rule, but you can refer to the literature\cite{linProjectedGradientMethods2007}.
Now, we can apply the PGD method to the subproblems in the ALS algorithm as they are constrained only by nonnegativity.
As the upper bound is not defined in the NMF problem, we can simply set the parameter $\mathbf{u}$ to be infinity.
When we describe the objective function of the subproblems as $f(W)$ and $f(H)$, its gradient term can be expressed as:
\begin{equation} 
  \begin{split} 
  \label{eq:pgd_nmf_gradient}
  \nabla f(W) = (V - WH) H^\top \\
  \nabla f(H) = W^\top (V - WH).
\end{split} 
\end{equation}

\subsection*{Nonnegative/Binary Matrix Factorization}
Nonnegative/binary matrix factorization (NBMF) is a variant of NMF that constrains the coefficient matrix $H$ to be binary and expressed as follows:
\begin{equation} 
  \begin{split} 
  \label{eq:NBMF}
  \min_{W, H} & \ ||V - WH||_F^2 \\
  \text{subject to} \quad & W \in \mathbb{R}^{+m \times k} \\
  & H \in \{0, 1\}^{k \times n}.
\end{split}  
\end{equation}

To approximate the solution of NBMF, we can similarly employ the ALS algorithm to iteratively update $W$ and $H$ as depicted in Algorithm \ref{alg:ALS_NBMF}.
The PGD method, described in the previous section, can solve the optimization of $W$ in each iteration.
The optimization of $H$ under the binary constraint becomes hard as the search space is discrete.
In practice, the update in the $H$ step is decomposed into $n$ independent binary optimization problems as:
\begin{equation} 
  \begin{split} 
  \label{eq:NBMF_H_decomposed}
  H^{t+1}_j := \argmin_{\mathbf{h} \in \{0, 1\}^{k}} ||V_j - W^{t+1} \mathbf{h}||_2^2,
\end{split}  
\end{equation}
where $X_j$ denotes the $j$-th column of matrix $X$.
In NBMF, even if we decompose the problem into subproblems \eqref{eq:NBMF_H_decomposed}, it is hard to solve them exactly.
To illustrate the difficulty, we can consider the case $n = 1$, which is equivalent to the well-known subset-sum problem.
The subset-sum problem is NP-hard, and no known polynomial-time algorithm exists to solve it \cite{garey1979computers}.
Consequently, solving problem \eqref{eq:NBMF_H_decomposed} is also NP-hard in general, and heuristic approaches are reasonable to obtain approximate solutions.
In the literature, quantum annealing has been proposed as a heuristic solver for NBMF\cite{omalleyNonnegativeBinaryMatrix2018b, goldenReverseAnnealingNonnegative2021b}.

\begin{algorithm}[tbp]
  \caption{Alternating Least Squares for NBMF}
  \label{alg:ALS_NBMF}
  \begin{algorithmic}
      \renewcommand{\algorithmicrequire}{\textbf{Input:}}
      \renewcommand{\algorithmicensure}{\textbf{Output:}}
      \Require{A nonnegative matrix $V \in \mathbb{R}^{+m \times n}$, the rank $k$}
      \Ensure{Nonnegative and binary matrices $W \in \mathbb{R}^{+m \times k}$ and $H \in \{0, 1\}^{k \times n}$}
      \State Initialize $W^1 \in \mathbb{R}^{+m \times k}$ and $H^1 \in \{0, 1\}^{k \times n}$.
      \While{not converged}
      \State  $W^{t+1}$ := $\argmin_{W \in \mathbb{R}^{+m \times k}} ||V - W H^t||_F^2$
      \State  $H^{t+1}$ := $\argmin_{H \in \{0, 1\}^{k \times n}} ||V - W^{t+1} H||_F^2$
      \State $t := t + 1$
      \EndWhile
  \end{algorithmic}
\end{algorithm}

\subsection*{Quantum Annealing}
Quantum annealing is a meta-heuristic inspired by quantum physics that leverages quantum fluctuations to escape local minima and find the global minimum of a given objective function.
Quantum annealing is specialized in solving quadratic unconstrained binary optimization (QUBO) problems, which can be expressed as:
\begin{equation}
  \begin{split}
    \text{minimize} & ~~~~ \mathbf{x}^\top Q \mathbf{x} \\
    \text{subject~to} & ~~~~ \mathbf{x} \in \{0,1\}^N,
  \end{split}
\end{equation}
where $Q$ is a coefficient matrix, $x$ represents a vector of a binary variable, and $N$ is the number of variables.
QUBO problems can be equivalently transformed into the problem of minimizing the energy of the Ising model, for which the Hamiltonian is expressed as:
\begin{equation}
  H_0(\mathbf{\sigma}) = - \sum_{i,j} J_{ij} \sigma_i \sigma_j - \sum_i h_i \sigma_i,
\end{equation}
where $\sigma_i$ is an Ising spin variable that takes either $1$ or $-1$ and $J_{ij}$ and $h_i$ are the coupling constant between Ising spins and the bias term, respectively.

To utilize quantum annealing, the objective function is encoded into an Ising Hamiltonian of a quantum system, and the ground state of the Hamiltonian corresponds to the optimal solution of the objective function.
For efficient exploration of low-energy states, the quantum annealing uses quantum fluctuations to escape local minima.
The quantum fluctuations are controlled by the transverse field, and the Hamiltonian of the quantum system is expressed as:
\begin{equation}
  \hat{H}(s) = - A \left(s(t)\right) \sum_i \hat{\sigma}_i^x + B \left(s(t)\right) \hat{H}_0,
\end{equation}
where $\hat{\sigma}_i^x$ represent the $x$-component of the Pauli matrices. 
The term $\hat{H}_0$ is the problem Hamiltonian, which encodes the Ising problem and is obtained by replacing the Ising spin variables with the $z$-component of the Pauli matrices, $\hat{\sigma}_i^z$.
The functions $A \left(s(t)\right)$ and $B \left(s(t)\right)$ are defined such that $A(0) \gg B(0)$ and $A(1) \ll B(1)$.
The parameter $0 \leq s(t) \leq 1$ is used to control the time evolution of the quantum system, which is called the annealing schedule.
In general, $s(t) = t / \tau$ is used, where $\tau$ is the total annealing time.
The standard quantum annealing schedule is set to decrease the transverse field linearly from $s(0) = 0$ to $s(\tau) = 1$.
At $s(0) = 0$, the system starts in a trivial ground state, making states uniformly superposition of all possible states.
By evolving the system over a sufficiently long time, the ground state of $\hat{H}_0$ can be obtained with a sufficiently high probability at $s(\tau) = 1$. 

To apply quantum annealing to NBMF, the problem \eqref{eq:NBMF_H_decomposed} can be encoded into a QUBO problem by defining the coefficient matrix $Q$ as follows:
\begin{equation}
  \label{eq:NBMF_QUBO}
  \begin{split}
    Q = W^\top W - 2 \text{diag} \left( V_j^\top W \right),
  \end{split}
\end{equation}
where $\text{diag}(\mathbf{v})$ denotes a diagonal matrix with the vector $\mathbf{v}$ on the diagonal.
Now, we can utilize quantum annealing to solve the QUBO problem \eqref{eq:NBMF_QUBO}.

Physical machines implementing quantum annealing is currently being developed by D-Wave Systems Inc., and they are publicly accessible \cite{johnsonQuantumAnnealingManufactured2011}. 
While D-Wave hardware faces challenges in terms of qubit count and connectivity, many real-world combinatorial optimization problems can be reduced to ground-state searches of Ising models, making quantum annealing an area of significant interest across various fields \cite{lucasIsingFormulationsMany2014b}.
In practice, various factors, such as noise, connectivity, and annealing time, limit the performance of D-Wave machines.
Various techniques have been proposed to overcome these limitations, such as reverse annealing (RA), a derivative procedure of quantum annealing that prioritizes search in a vicinity under given initial states.

\subsection*{Reverse Annealing}
RA is a technique focused on local search \cite{chancellorModernizingQuantumAnnealing2017b, ohkuwaReverseAnnealingFully2018b}. 
When $ t = 0 $, a classical initial state is prepared with $ s(0) = 1 $, setting the transverse field term to zero. 
Then, the strength of the transverse field term gradually increases until it reaches a predetermined switching point $ s(t_{\text{inv}}) = s_{\text{target}} $. 
We refer to the reversed switching point $ 1 - s_{\text{target}} $ as the reversal distance.
Afterward, as in the standard quantum annealing (i.e., FA), the transverse field is weakened, allowing the system to find a lower-energy state close to the initial state. 
In this way, RA can refine the initial states and possibly find a better solution than FA alone.

One of the known limitations of D-Wave machines is that external influences can prevent ideal quantum annealing execution, resulting in suboptimal solutions with less-than-ideal probabilities. 
However, since D-Wave machines can output relatively low-energy solutions within microseconds, they are positioned similarly to heuristic methods like simulated annealing \cite{kirkpatrickOptimizationSimulatedAnnealing1983a}. 
D-Wave machines offer a ``pause'' feature, which halts the annealing parameter at a certain magnitude $ s = s^* $ over an extended period. 
During RA on D-Wave machines, the pause function can be used at the switching point $ s(t_{\text{inv}}) = s_{\text{target}} $ to fix the transverse field, enabling solution search through thermal bath relaxation, a widely adopted technique.

To calibrate the annealing schedule, the annealing time $ \tau $ and the switching point $ s_{\text{target}} $ must be adjusted as they are dominant factors in the performance of RA.
Especially, the annealing distance defined by the switching point is crucial for the performance of RA as it controls the strength of the quantum fluctuations and how far the system can traverse from the initial state.
If we set the reversal distance to high, the system will completely forget the initial state and run as the FA equivalently.
On the other hand, if the reversal distance is too short, the system will not have enough fluctuation to escape from the initial state, and the performance of RA will be degraded.
The preceding study calibrated the reversal distance by evaluating the probabilities to escape from the initial state and reach a better solution\cite{goldenReverseAnnealingNonnegative2021b}.
The length of pausing during the annealing schedule did not have much impact on the performance of RA.
This study uses the same annealing schedule for comparison with the preceding study.

\subsection*{Relaxation-Assisted Reverse Annealing}
In RA, the performance is highly dependent on the initial states.
In the preceding study, the states in each iteration in the ALS algorithm were utilized as the initial configuration for RA.
This study proposes an improved strategy integrating RA with a classical linear programming relaxation technique.
We apply the relaxation to the problem \eqref{eq:NBMF_H_decomposed} and obtain the following relaxed problem:
\begin{equation} 
  \begin{split} 
  \label{eq:NBMF_relaxed}
  H^{t+1}_j := \argmin_{\mathbf{h} \in [0, 1]^{k}} ||V_j - W^{t+1} \mathbf{h}||_2^2,
\end{split}  
\end{equation}
where $[0, 1]$ denotes the set of real numbers between $0$ and $1$.
The relaxed problem \eqref{eq:NBMF_relaxed} can be solved by the PGD method with brief modification.
We only need to change the projection function \eqref{eq:projection} in the PGD method and simply set the upper bound parameter $\mathbf{u}$ as a vector of ones.

The relaxed solutions can be obtained by the PGD method and we use them for initial configuration for RA.
As the relaxed solution is continuous, we need to map the relaxed solution to binary.
There are several methods to map the continuous solution to binary, such as rounding or random sampling.
In this study, we use the simple rounding method to map the relaxed solution to binary by the following rule:
\begin{equation} \begin{split} 
  \label{eq:rounding}
  h_i = \begin{cases}
    1 & \text{if} ~ h_i \geq 0.5 \\
    0 & \text{otherwise}.
  \end{cases}
\end{split}  \end{equation}
Although the rounding method does not always yield the optimal solution, it approximates the original problem well.
This rounded solution can then serve as the initial configuration for RA, which is expected to further refine the approximation, guiding it towards an optimal or near-optimal solution.

\section*{Results}
\subsection*{Practical Performance Evaluation on Facial Image Datasets}
We evaluate the performance of the proposed method on facial image datasets, which is used in the literature for NMF and NBMF \cite{leeLearningPartsObjects1999,omalleyNonnegativeBinaryMatrix2018b,goldenReverseAnnealingNonnegative2021b}.
We targeted 200 images, each with a resolution of $19 \time 19$ pixels, from the publicly available MIT face image dataset \cite{cbcl_face_database_2000}.
To convert the image dataset to a matrix $V$, we vectorized each image and stacked them into a matrix $V \in \mathbb{R}^{+361 \times 200}$.
We set the rank of the factorization to $k = 35$ and initialized the basis matrix $W$ and the coefficient matrix $H$ randomly.
We applied the ALS algorithm to factorize the matrix $V$ into $W$ and $H$.
The optimization of $W$ was solved by the PGD method, and the optimization of $H$ was solved by the proposed relaxation-assisted RA method.
To evaluate the performance of the proposed method, we compared the different methods used in the literature as shown in Table \ref{tab:methods}.

\begin{table}[tbp]
  \caption{
    \textbf{Method overview}
  }
  \centering
  \small
  \begin{tabular}{llllc}
    \hline
    Method & Description & Previous $H$ & Initial states for RA & Literature \\
    \hline
    Exact & Exact algorithm & Unused & -- & -- \\
    PGD & PGD and rounding & Used & -- & -- \\
    FA & FA only & Unused & -- & \cite{omalleyNonnegativeBinaryMatrix2018b} \\
    RA & RA without assist & Used & The previous state & \cite{goldenReverseAnnealingNonnegative2021b} \\
    RA+FA & RA initialized by FA & Unused & FA & -- \\
    \bf{RA+PGD} & Relaxation-assisted RA & Used & Relaxation & Proposed \\
    \hline
  \end{tabular}
  \label{tab:methods}
\end{table}

As devices to perform FA and RA, we used the D-Wave 2000Q, a machine publicly available from D-Wave. 
For the annealing schedule and execution frequency, we used previously demonstrated effective parameters in the preceding research \cite{goldenReverseAnnealingNonnegative2021b}.
The annealing time for FA was set to 20 microseconds.
We implemented a total annealing time of 30 microseconds for RA, including a 10-microsecond pause. 
The number of executions for FA and RA were set to 1,000 and 240, respectively.
This setting was determined to nearly equalize the total execution time of FA and RA.
The problem must be mapped onto the physical processor structure called the quantum processing unit (QPU) to implement the Ising model on D-Wave's QA machine. 
In this case, we used minor embedding to implement it as a complete graph.
The majority vote method was also applied for solution recovery if qubit chain breaks occurred.
The computations for the classical computation were performed using an Intel Xeon Gold 6130 CPU with 141 GB of RAM.
The PGD algorithm was implemented in Python 3.8.8 by referring to Matlab implementations in the literature \cite{linProjectedGradientMethods2007}.
To compare the performance of the proposed method, we evaluated the optimal cases using the commercial solver Gurobi Optimizer 9.1.2, widely recognized as the industry standard for solving complex optimization problems involving integer programming.

\begin{figure}[ht]
  \centering
  \includegraphics[width=0.7\textwidth]{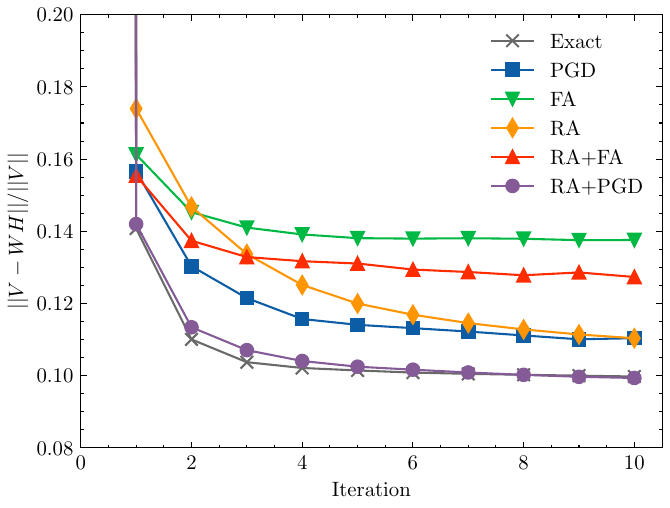}
  \caption{
    \textbf{Squared error at each iteration of the ALS method.}
    The plot shows the squared error at each iteration of the ALS method by using methods shown in Table \ref{tab:methods}. 
  }
  \label{fig:error_vs_iter}
\end{figure}

The evaluation of squared error at each iteration of the ALS method is shown in Fig. \ref{fig:error_vs_iter}. 
The FA-only approach exhibited lower learning performance than the PGD and rounding-based methods. 
Additionally, even when RA was applied with FA, it did not surpass these methods.
In RA using the previous solution, the convergence error was superior to that of FA.
Moreover, relaxation-assisted RA achieved performance closely matching the exact cases.
We recognized that the relaxation-assisted RA method was the most effective compared to other configurations for RA.

\begin{figure}[ht]
  \centering
  \includegraphics[width=0.7\textwidth]{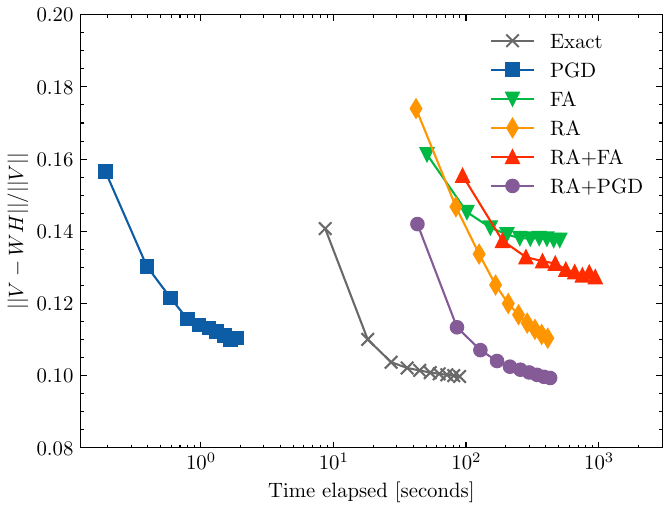}
  \caption{
    \textbf{Squared error over elapsed time.}
    The plot shows the squared error at each iteration of the ALS method by using methods shown in Table \ref{tab:methods}. 
  }
  \label{fig:error_vs_time}
\end{figure}

Next, we evaluated the computational performance of each method by measuring the computation time required to obtain an output to the problem \eqref{eq:NBMF_H_decomposed}.
The total QPU access time was used for FA and RA, including the annealing time $\tau$ and pre-processing and post-processing time.
The learning performance of each method over elapsed time is shown in Figure \ref{fig:error_vs_time}.
Based on these results, RA with the relaxation assist was confirmed to be faster than RA with an initial state from QA from a computation time perspective.
However, in this experimental setup, the Gurobi Optimizer method was even faster and more effective than any RA-based method.
This is likely because the RA process is more time-consuming than preparing the initial state, and exact methods are still more efficient for this small-scale problem.

RA with an initial state estimated through relaxation outperformed conventional RA configuration.
It is natural due to the high-precision optimization of $H$ in each iteration of the ALS method, as represented by Equation \eqref{eq:NBMF_H_decomposed}.
Since the performance of RA depends on the initial state, the approximated solution through relaxation is expected to be advantageous for RA.
To clarify the amount of performance improvement by RA, we analyzed the Hamming distance to the optimal solution at each iteration of the ALS method.
For two solutions $ \mathbf{x} $ and $ \mathbf{x}' $, the Hamming distance $ d(\mathbf{x}, \mathbf{x}') $ is defined as:
\begin{equation}
  d(\mathbf{x}, \mathbf{x}') = \sum_{i} |x_i - x_i'|.
\end{equation}
The Hamming distance represents the number of variables that differ between the two values.
As long as we do not consider problems where multiple optimal solutions do not exist, calculating the Hamming distance between the result and the optimal solution provides a measure of proximity to the optimal solution.

\begin{figure}[ht]
  \centering
  \includegraphics[width=1\textwidth]{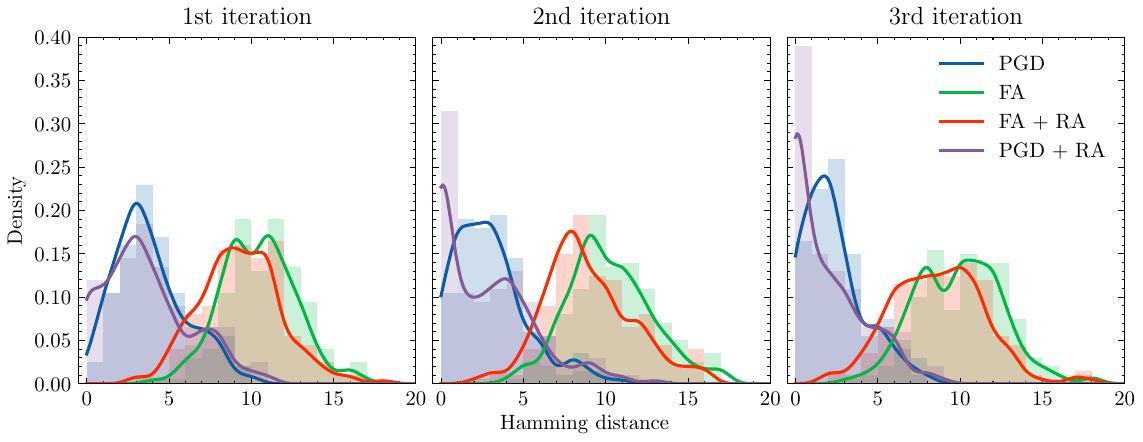}
  \caption{
    \textbf{Hamming distance from the optimal solution at each iteration of the ALS method.} At each iteration, We analyzed the output for the 200 subproblems represented by Equation \eqref{eq:NBMF_H_decomposed}. The solid line represents the probability density obtained by kernel density estimation.
  }
  \label{fig:Hamming_distance}
\end{figure}

For the dataset used in the previous section, we examine the proximity at each iteration of the ALS method.
We illustrate the result in Figure \ref{fig:Hamming_distance}.
Compared to FA, PGD was confirmed to output solutions closer to the optimal solution.
Furthermore, it was found that the percentage of optimal solutions obtained increases through relaxation-assisted RA as the iteration progresses. 
In contrast, FA, which cannot utilize the previous $ H $'s information, did not show improvements in Hamming distance and kept the same level of optimization performance.
After all, the relaxation approach was confirmed to be advantageous, as it could provide a better initial state for RA.

Some say that initialization is the most beneficial factor in the performance of RA rather than the annealing process itself.
In our experiments, the PGD and the rounding-based method itself did not reach the optimal cases, but RA was successful in refining the solution and achieving almost the optimal solution.
This result suggests that RA can effectively refine the solution obtained by classical methods if the initial state is close to the optimal solution but hard to reach.
To investigate the potential of relaxation-assisted RA, we need to explore how RA can refine the initial state obtained by the relaxation method and how the relaxation method can be improved to provide a better initial state for RA.

\begin{figure}[ht]
  \centering
  \includegraphics[width=0.7\textwidth]{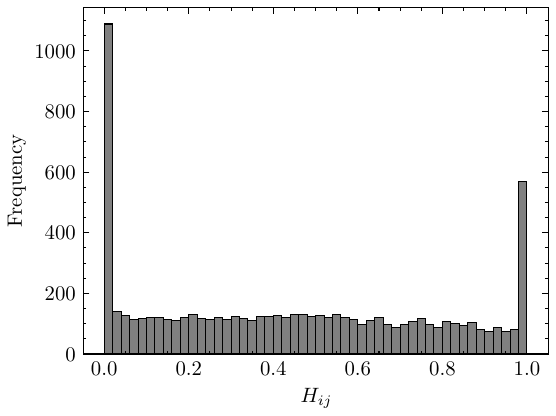}
  \caption{
    \textbf{Distribution of elements in $H$.} We analyzed the matrix $H$ obtained by PGD and the rounding-based method after a single iteration of the ALS method. The histogram represents the distribution of the elements in $H$.
  }
  \label{fig:H_Distribution}
\end{figure}

Ultimately, we analyzed the quality of relaxed solutions obtained by the PGD methods.
The distribution of the elements in $H$ at the first step in ALS is illustrated in Figure \ref{fig:H_Distribution}.
The distribution of the elements in $H$ was concentrated around $0$ and $1$ and strongly contrasted between them.
This result indicates that the rounding procedure is unlikely to significantly degrade the quality of the solution.
Thus, the initial states obtained by the relaxation and rounding method are believed to be a good approximation of the original problem and, in the next subsection, we investigate further on the quality of the relaxation-based initialization for RA.
The superior performance of the relaxation-based approximation method can likely be attributed to the characteristics of the facial dataset used in this study.
Facial datasets tend to have relatively well-defined and learnable features, making it easier to capture and represent these features accurately in practice.
As a result, factorizing the original matrix becomes less challenging, allowing the relaxation-based method to perform effectively under these conditions.
For more complex problems, such as those involving randomly generated datasets, the relaxation-based approximation may not perform as effectively as it does in this study. 
However, datasets encountered in nature or in practical applications are often structured and contain patterns or relationships that can be leveraged. 
The relaxation-based approximation, followed by refinement through RA, can be effective in such cases. 

\subsection*{Estimation Accuracy of Relaxation-Based Initialization}
In this subsection, we assess the estimation performance of the relaxation-based initialization method.
In facial image datasets, the relaxation-based initialization method was found to yield a better initial state for RA, offering closer proximity to the optimal solution.
As shown in Fig. \ref{fig:H_Distribution}, the elements of the relaxed solution are concentrated around $0$ and $1$, a property that we believe facilitates a good approximation of the original problem.
To identify the limitations of the relaxation-based initialization method, we examine how its performance generalizes to other NBMF problems.
Specifically, we prepare randomly generated datasets and evaluate the performance in terms of the proximity depending on the rank of the factorization, which corresponds to the number of variables.

\begin{figure}[ht]
  \centering
  \includegraphics[width=0.95\textwidth]{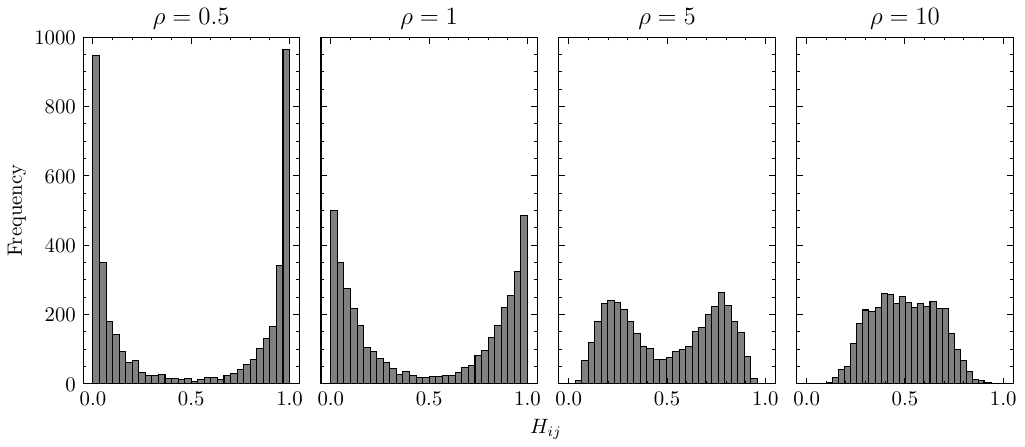}
  \caption{
    \textbf{Example distributions of the randomly generated relaxed solution.} Each plot illustrates the distribution of the elements in the relaxed solution randomly generated by our procedure, with shape parameter $\rho$ set to 0.5, 12, 2, and 10. When $\rho$ is small, the solution is concentrated around 0 and 1; as $\rho$ increases, the distribution becomes more Gaussian-like.
  }
  \label{fig:gamma_distribution}
\end{figure}

To control the distribution of the relaxed solution, we synthetically generate the coefficient matrix 
$H$ using the following procedure:
\begin{enumerate}
  \item  $A \sampling^{nk} \gamma(\rho, \theta)$
  \item  $A \leftarrow A / \max (A)$
  \item  $B \leftarrow \mathbf{1} - A[1:\lfloor nk/2 \rfloor], ~~ C \leftarrow A[\lfloor nk/2 \rfloor + 1:nk]$
  \item  Randomly select elements from $B$ and $C$ without replacement and store them in $H$,
\end{enumerate} 
where $X \sampling^{N} \gamma(\rho, \theta)$ denotes sampling $N$ elements from the gamma distribution and storing them in the array $X$.
The gamma distribution is defined as:
\begin{equation*}
  f(x) = \frac{1}{\Gamma (\rho) \theta ^\rho} x^{\rho-1} e ^{-x/\theta},
\end{equation*}
where $\rho$ and $\theta$ are the shape parameters, and $\theta = 1$ is used in this study.
The array $\mathbf{1}$ is a vector of length $\lfloor nk/2 \rfloor$ whose elements are all set to $1$.
This procedure allows us to control the distribution of the relaxed solution by varying the shape parameter $\rho$ of the gamma distribution.
An example of the relaxed solution's distribution is shown in Fig. \ref{fig:gamma_distribution}.
The relaxed solution was concentrated around $0$ and $1$ when $\rho$ was small, and the distribution became close to the Gaussian-like distribution as $\rho$ increased.
Finally, we construct the basis matrix $W$ by sampling from the gamma distribution with the parameters $\rho = 1$ and $\theta = 1$.
The original data $V$ is then obtained by the product of $W$ and $H$.
In this manner, the dataset is generated to mimic the facial image dataset used in the previous section, while the shape parameter $\rho$ governs the inherent contrast in the relaxed solution.
In typical NMF tasks, it is recommended to choose the number of features $k$ so that $k$ to satisfy $k(n+m) < nm$ to avoid overfitting.
Therefore, we set $m$ to $2nk / (n - 2k)$ to keep the problem complexity roughly consistent across different values of $n$ and $k$.

\begin{figure}[ht]
  \centering
  \begin{minipage}{0.49\hsize}
    \centering
    \includegraphics[width=1\textwidth]{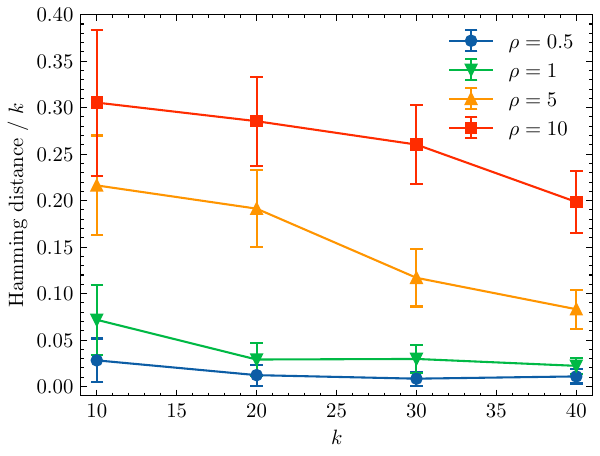}
    \captionsetup{labelformat=empty, justification=centering} 
    \caption{(a)}
    
    \label{fig:gamma_error_rate}
  \end{minipage}
  \begin{minipage}{0.49\hsize}
    \centering
    \includegraphics[width=1\textwidth]{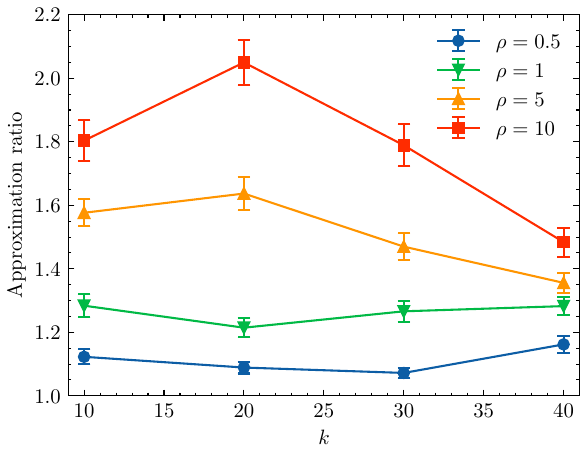}
    \captionsetup{labelformat=empty, justification=centering} 
    \caption{
      (b)
    }
    \label{fig:gamma_error_rate_k}
  \end{minipage}
  \label{fig:performance_synthetic}
  \caption{
    \textbf{Estimation performance of the relaxation-based initialization method.} 
    Each panel shows performance results for the relaxation-based initialization method. In (a), the average of Hamming distance from the optimal solution is shown, while in (b), the average of approximation ratio is plotted. Error bars indicate one standard error of the mean. The shape parameter $\rho$ is set to 0.5, 1, 2, and 10. For $k = 10, 20, 30$, there are 110 samples; for $k = 40$, 104 samples are used, as some took extremely long to solve and were omitted.
  }
\end{figure}

Using this synthetic dataset, we evaluate the estimation performance of the relaxation-based initialization method.
To measure proximity to the optimal solution, we use the Gurobi Optimizer to find the optimal solution.
To prevent the solver from stalling in rare cases, we set a one-minute time limit, which is almost sufficient to find the optimal solution for the problem sizes in this study.
We fixed the number of data points, $n$, at 110 and analyzed the approximation quality for each sample derived by solving the problem \eqref{eq:NBMF_H_decomposed}.
Figure \ref{fig:gamma_error_rate} illustrates the estimation performance of the relaxation-based initialization method.
The results confirm that when the shape parameter $\rho$ is small and the relaxed solution is concentrated near $0$ and $1$, the relaxation-based initialization method offers superior proximity to the optimal solution.
As $\rho$ increased, proximity to the optimal solution declined.
Regarding the number of features $k$, proximity to the optimal solution tends to improve as $k$ increased.
A similar trend was observed in the approximation ratio, defined as the ratio of the optimal objective value to the one obtained by the relaxation-based initialization method.
From these results, we can conclude that the relaxation-based initialization method is effective for problems where the relaxed solution is concentrated around $0$ and $1$.
Furthermore, for a larger number of features, the relaxation-based initialization method can offer a closer approximation to the optimal solution without degrading estimation quality.
This advantage explains why the relaxation-based initialization method performs well on the facial image dataset in the previous section, where strong contrast in feature weights leads to a relaxed solution concentrated near 0 and 1 as depicted in Fig. \ref{fig:H_Distribution}.
This result suggests that when applyng quantum annealing to certain problems, one can efficiently obtain a high-quality initial guess for RA, and  potentially improving the overall performance of quantum annealing.

\section*{Discussion}
In this study, we proposed a hybrid approach combining relaxation-based initialization and RA in NBMF.
For the relaxation, by extending the PGD method used in traditional NMF, we demonstrated that highly accurate estimated solutions could be obtained for original problems.
The proposed method was evaluated on facial image datasets, and the results showed that relaxation-assisted RA outperformed conventional RA methods and achieved performance close to the optimal cases.
Due to their closeness to the optimal states, the relaxation-based initializations proved advantageous for RA by offering initial states that were near the optimal solution in terms of Hamming distance.
We further analyzed the quality of the relaxed solutions obtained via relaxation-based initialization and confirmed its effectiveness for the facial image dataset used in this study.
In addition, we evaluated the estimation performance of the relaxation-based initialization method on synthetic datasets, confirming its effectiveness when the relaxed solution is concentrated around $0$ and $1$.
However, we found that the performance deteriorated when the relaxed solutions adopted a more Gaussian-like distribution.
Nevertheless,  as the number of features $k$ increased, the relaxation-based initialization method continued to offer a better approximation to the optimal solution.

Since real-world datasets, such as facial images, often exhibit significant bias in feature weights and have contrast, we believe the proposed method is effective for practical applications. 
However, a drawback of the linear programming relaxation estimation method is that the approximation quality tends to deteriorate if the relaxed solution's distribution has few values near $0$ or $1$.
The maximum independent set problem is one example of a problem where relaxation-based initialization seems ineffective, and relaxed solutions tend to concentrate at $0.5$.
The superiority of our method over other general optimization algorithms in addressing such challenging problems remains uncertain.
We believe it is crucial to explore alternative initial approximation methods for RA and investigate how the choice of initial states affects final solution quality.

This study is likely to broaden the applicability of RA not only in NBMF but also in other combinatorial optimization problems. 
The results demonstrate that efficient preparation of initial states can achieve performance unattainable by FA alone.
RA could be an effective approach for optimization algorithms that update iteratively and for problems where highly accurate approximate solutions are achievable using the previous solution as information.
For example, in tackling problems involving complicating variables that take real values, Benders' decomposition can be an effective approach\cite{bendersPartitioningProceduresSolving1962,rahmanianiBendersDecompositionAlgorithm2017}.
This method decomposes the original problem into subproblems by separating the complicating variables from the rest. Each subproblem is then solved iteratively, with solutions feeding back into the main problem, allowing for an efficient convergence toward the optimal solution.
A relaxation-assisted RA approach can be well-suited to updating and refining the solutions for subproblems involving binary variables.

\section*{Acknowledgments}
This study was financially supported by programs for bridging the gap between R\&D and IDeal society (Society 5.0) and Generating Economic and social value (BRIDGE) and Cross-ministerial Strategic Innovation Promotion Program (SIP) from the Cabinet Office.

\section*{Author contributions statement}
\addcontentsline{toc}{section}{Author contributions statement}
R.H. conceived of the presented idea and performed the experiments.
M.O. and K.T. verified the analytical methods and supervised the findings of this work. 
All authors discussed the results and contributed to the final manuscript.

\section*{Additional information}

\begin{itemize}
  \item \textbf{Competing interests}: The authors declare no competing interests. 
  \item\textbf{Correspondence} should be addressed to R.H.
  \item\textbf{Data availability}: The facial image dataset used in this study is publicly available from MIT at http://cbcl.mit.edu/cbcl/software-datasets/FaceData.html. Other datasets supporting the findings of this study are available upon reasonable request from the corresponding author.
\end{itemize}

\nolinenumbers


\end{document}